\documentclass{iopart}

\usepackage[]{graphicx}

\begin{document}

\JPA

\title[Stochastic dynamics and thermodynamics around a metastable state]{Stochastic dynamics and thermodynamics around a metastable state based on the linear Dean-Kawasaki equation}

\author {Hiroshi Frusawa\footnote[1]{e-mail: frusawa.hiroshi@kochi-tech.ac.jp}}
\address{Laboratory of Statistical Physics, Kochi University of Technology, Tosa-Yamada, Kochi 782-8502, Japan.}

\begin{abstract}
The Dean-Kawasaki equation forms the basis of the stochastic density functional theory (DFT).
Here it is demonstrated that the Dean-Kawasaki equation can be directly linearized in the first approximation of the driving force due to the free energy functional $F[\rho] $ of an instantaneous density distribution $\rho$, when we consider small density fluctuations around a metastable state whose density distribution $\rho^*$ is determined by the stationary equation $\delta F[\rho]/\delta \rho|_{\rho=\rho^*}=\mu$ with $\mu$ denoting the chemical potential.
Our main results regarding the linear Dean-Kawasaki equation are threefold.
First, (i) the corresponding stochastic thermodynamics has been formulated, showing that the heat dissipated into the reservoir is negligible on average.
Next, (ii) we have developed a field theoretic treatment combined with the equilibrium DFT,  giving an approximate form of $F[\rho]$ that is related to the equilibrium free energy functional.
Accordingly, (iii) the linear Dean-Kawasaki equation, which has been reduced to a tractable form expressed by the direct correlation function, allows us to compare the stochastic dynamics around metastable and equilibrium states, particularly in the Percus-Yevick hard sphere fluids; we have found that the metastable density is larger and the effective diffusion constant in the metastable state is smaller than the equilibrium ones in repulsive fluids. 
\end{abstract}

\section{Introduction}

The equilibrium density functional theory (DFT) has been most successful in the study of inhomogeneous classical fluids, precisely predicting their thermodynamic and structural properties in equilibrium [1-7].
The dynamical DFT (DDFT) extends the equilibrium DFT to describe an overdamped dynamics [8-15].
An overdamped dynamics is typically used to model soft matter systems, such as colloidal suspensions and supercooled liquids, that are characteristic of the slow dynamics on time scales larger than the relaxation time of momenta. 
In the supercooled liquids, the non-vibrational diffusive motions take much more time than collisions due to the strong correlation of the constituent particles.
Accordingly, it is plausible to assume that the momentum and the energy flow much faster via collisions through the system than the slowly decaying number density.
The colloidal suspensions, on the other hand, consist of micrometer-sized particles suspended in a molecular solvent.
These systems have the advantage that the solvent stays equilibrated in any situation and acts as a bath at the given temperature. 
It follows that the solvent microscopic degrees of freedom can be traced out to allow the colloids to be treated as the overdamped Brownian particles.

The DDFT framework has thus been shown to successfully model a wide variety of fluid dynamics phenomena, including crystallization [11] and steady state structures in driven suspensions [12-15].
There is also a limitation that the DDFT is constituted by the deterministic equation and is not applicable to processes governed by fluctuations.
One way of overcoming the limitation of the DDFT is to incorporate the influence of the fluctuating solvents into the noise term, which will be referred to as the stochastic DFT [16-46].
The stochastic DFT is  based on the so-called Dean-Kawasaki (DK) equation that inherently describes the evolution of the instantaneous microscopic density field [16-21].
For almost two decades, it has been an open question as to whether the DK equation containing the microscopic information can represent the dynamical law of observable quantities;
particularly in terms of the coarse-graining procedure, the long-standing problem of the DK description has been investigated in the literature [20-26, 38].
Furthermore, the two-body interactions in the original DK equation are expressed by the bare interaction potential, instead of by the direct correlation function whose use has been validated in the DDFT, indicating that the DK equation is applicable to soft core systems [31] but is basically irrelevant to harshly repulsive ones.
A variety of coarse-graining procedures has therefore been explored not only for smoothing the interaction potential appropriately [17, 18], but also for smearing the microscopic density [21-26].
Nevertheless, the validity of the DK equation in harshly repulsive systems remains to be settled [26, 28-30, 38].

While addressing the fundamental issue, the stochastic DFT has been used as one of the most powerful tools for describing slowly fluctuating and/or intermittent phenomena such as glassy dynamics [27-30], nucleation or pattern formation of colloidal particles [31, 32], stochastic thermodynamics of colloidal suspensions [33], dielectric relaxation of Brownian dipoles [34, 35], and even tumor growth [36].
It has been recently found that linear stochastic equation of density fluctuations is much of practical use [37-46]:
the density fluctuations of fluids near equilibrium are surprisingly well described by model-B dynamics of a Gaussian field theory whose effective quadratic Hamiltonian for the density fluctuation field is constructed to yield the exact form of the static density-density correlation function [37, 38].
Since the DK equation can be reduced to the linear stochastic equation under some approximations, such kind of Gaussian theory for fluctuations may be categorized as a stochastic DFT.
Actually, the linear stochastic DFT has also been found relevant to investigate out-of-equilibrium phenomena, including not only the evaluation of the work performed in active and driven particle systems, but also the formulations of the full Onsager theory of electrolyte conductivity [40], the instantaneous stress tensor [42] for interacting Brownian particle, and the long-ranged fluctuation-induced (Casimir) forces in active [44] and thermal matter [45-46]. 

We aim to develop an alternative to the above linear stochastic DFT that applies to the stochastic dynamics and thermodynamics around a metastable state defined below, not near equilibrium.
Our focus is on small fluctuations in density field around the metastable state, so that the original DK equation can be directly linearized with its bare interaction potential altered to a tractable form, without entering into the details of the coarse-graining procedure.

This paper proceeds as follows:
In section 2, we describe the formal background and our scheme for revising the linear stochastic DFT [37-46].
In section 3, the linear stochastic density functional equation is directly derived from the DK equation around a metastable state defined therein.
Section 4 investigates the stochastic thermodynamics [33] around a metastable state using the stochastic DFT formulated in section 3, which validates that the heat dissipated due to small fluctuations around metastable density distribution is negligible.
To go beyond the general formulations presented in sections 3 and 4, we need to obtain a tractable form of the free energy functional $F[\rho]$ of a given density profile $\rho$ (see section 2 for the detailed definition of $F[\rho]$).
In section 5, we demonstrate that a field theoretic approach combined with the equilibrium DFT leads to an approximate form of $F[\rho]$
that is expressed by the direct correlation function but has an additional term to the equilibrium free energy functional used in the equilibrium DFT.
In section 6, we clarify the change of relaxation dynamics due to the metastability not in equilibrium, using the additional contribution of $F[\rho]$ to the equilibrium free energy functional evaluated in section 5.

\section{Formal background and our scheme}

We consider $N$-particles interacting via potential $v(|{\bf x}-{\bf y}|)$ each of which moves in the external potential $k_BTJ({\bf x})$ surrounded by a heat bath at temperature $T$.
We can relate a coarse-grained density $\rho({\bf x},t)$ to the microscopic density $\hat{\rho}({\bf x},t)$ via introducing the distribution functional $P[\rho,t]$ of a given density profile [20]:
\begin{equation}
P[\rho,t]=\left<
\prod_{{\bf x}}\delta\left[
\rho({\bf x},t)-\hat{\rho}({\bf x},t)
\right]
\right>,
\label{distribution}
\end{equation}
where $\left<\mathcal{O}\right>$ denotes the averaging operation for $\mathcal{O}$ in the overdamped dynamics (see the literature [20-22, 26] for the explicit representation of averaging), and the microscopic density $\hat{\rho}({\bf x},t)$ of the $N$-particle system is given by
\begin{equation}
\hat{\rho}({\bf x},t)=\sum_{i=1}^N\delta[{\bf x}-{\bf x}_i(t)],
\label{microscopic}
\end{equation}
using the position ${\bf x}_i(t)$ of the $i$-th particle at time $t$.
The distribution functional $P[\rho,t]$ satisfies the Fokker-Planck equation such that [20, 21, 38]
\begin{eqnarray}
\partial_t P[\rho,t]=-\int d{\bf x}
\frac{\delta}{\delta\rho}\nabla\cdot
D\rho\,\nabla
\left[
\frac{\delta}{\delta\rho}+\frac{\delta\left(\beta F[\rho]\right)}{\delta\rho}
\right]
 P[\rho,t],
\label{fp}
\end{eqnarray}
where $F[\rho]$ corresponds to the free energy functional of a given density profile that reads [19-21, 38]
\begin{eqnarray}
\beta F[\rho]
&=&\frac{1}{2}\int d{\bf x}\int d{\bf y}\,
\mathcal{G}_{\rho}({\bf x},{\bf y})\beta v(|{\bf x}-{\bf y}|)
+\int d{\bf x}\rho({\bf x},t)J({\bf x})\nonumber\\
&&\qquad\qquad\qquad +\int d{\bf x}\,\left\{
\rho({\bf x},t)\ln\rho({\bf x},t)
-\rho({\bf x},t)
\right\},
\label{f-rho1}
\end{eqnarray}
with $\beta$ denoting the inverse of thermal energy $k_BT$ and $\mathcal{G}_{\rho}$ representing an instantaneous correlation function defined by
\begin{eqnarray}
\mathcal{G}_{\rho}({\bf x},{\bf y})=\rho({\bf x},t)\rho({\bf y},t)
-\rho({\bf x},t)\delta(|{\bf x}-{\bf y}|).
\label{bare_correlation}
\end{eqnarray}
The Fokker-Planck equation (\ref{fp}) is equivalent to the Langevin equation (the so-called DK equation) of the following form [16-21]:
\begin{equation}
\partial_t\rho=\nabla\cdot D\rho\,\nabla\frac{\delta(\beta F[\rho)]}{\delta\rho}+\zeta[\rho,{\bf \eta}],
\label{sdft0}
\end{equation}
wehre $\zeta[\rho,{\bf \eta}]$ can be expressed as [38]
\begin{equation}
\zeta[\rho,{\bf \eta}]=\nabla\cdot\sqrt{2D\rho({\bf x},t)}{\bf \eta}({\bf x},t)
\end{equation}
using the bare diffusion constant $D$ and the vectorial white noise field ${\bf \eta}({\bf x},t)$ defined by the correlation $\left<\eta_l({\bf x},t)\eta_m({\bf y},t')\right>=\delta_{lm}\delta({\bf x}-{\bf y})\delta(t-t')$.
It follows that the spatio-temporal averaging of $\zeta$ reads [38]
\begin{equation}
\fl
\left<\,\zeta[\rho({\bf x},t),{\bf \eta}({\bf x},t)]\>
\zeta[\rho({\bf y},t'),{\bf \eta}({\bf y},t')]
\,\right>
=-2D\delta(t-t')\nabla_{{\bf x}}\cdot\rho({\bf x},t)\nabla_{{\bf x}}\delta({\bf x}-{\bf y}).
\end{equation}

The strategy for reformulating the DK equation (\ref{sdft0}) is twofold:
(i) we will expand $\delta F[\rho]/\delta\rho$ causing the driving force in eq. (\ref{sdft0}) around a stationary condition representing the metastable and (ii) we will obtain an approximate but tractable form of $F[\rho]$ using a field theoretic treatment of $P_{\mathrm{eq}}[\rho]$ that is obtained from the steady condition $\partial_tP_{\mathrm{eq}}[\rho]=0$ in eq. (\ref{fp}) [20]:
\begin{eqnarray}
P_{\mathrm{eq}}[\rho]
&=&\left<\prod_{{\bf x}}\delta\left[
\rho({\bf x},t)-\hat{\rho}({\bf x},t)
\right]\right>_{\mathrm{eq}}\nonumber\\
&=&\frac{e^{-\beta F[\rho]}}{\int \mathcal{D}\rho\,e^{-\beta F[\rho]}}\delta\left(
\int d{\bf x}\,\rho({\bf x},t)-N
\right),
\label{peq-answer}
\end{eqnarray}
where the averaging operation $\left<\mathcal{O}\right>_{\mathrm{eq}}$ in equilibrium now follows the canonical ensemble (see Appendix A for the details).

Before proceeding, Table 1 lists similar symbols having different meanings of densities, distribution functionals, and free energy functionals.

\begin{table}
\caption{Similar symbols having different meanings.}
\centering
\begin{tabular}{c|l}
Symbol & Meaning\\
\hline \hline
$\hat{\rho}$ & Microscopic density defined by eq. (2).\\[3pt]
$\rho$ & Coarse-grained density having the identical distribution of $\hat{\rho}$.\\[3pt]
$\rho^*$ & Metastable density that will be defined by eq. (10).\\[3pt]
$\phi$ & Density difference defined by $\phi=\rho-\rho^*$.\\[3pt]
$\rho_{\mathrm{eq}}$ or $\left<\hat{\rho}\right>_{\mathrm{eq}}$ & Equilibrium density in the presence of an external field\\
& that will be defined by eq. (28).\\[3pt]
$\overline{\rho}$ & Smeared density defined by $\overline{\rho}=N/V$ using the system volume $V$.\\[3pt]
$P[\rho,t]$ & Time-evolving probability of a given density profile $\rho$.\\[3pt]
$P_{\mathrm{eq}}[\rho]$ & Time-invariant probability in a steady state of a given density profile $\rho$.\\[3pt]
$F[\rho]$ & Free energy functional of a given density profile $\rho$ defined by eq. (4).\\[3pt]
$\mathcal{F}_{\mathrm{HK}}[\rho]$ & Equilibrium free energy functional\\
& that will be defined by eqs. (29) and (30). 
\end{tabular}
\end{table}

\section{Stochastic DFT around a metastable state: general formulation}

As a metastable state, we would like to consider the density distribution, $\{\rho^*({\bf x},t)\}$, that is obtained from the stationary equation (\ref{fp}) as follows:
\begin{equation}
\left.
\frac{\delta (\beta F[\rho])}{\delta\rho}\right|_{\rho=\rho^*}=\mu,
\label{stationary1}
\end{equation}
where a constant, $\mu$, denotes the Lagrangian multiplier for ensuring that $\int d{\bf x}\,\rho({\bf x},t)-N$ and is equal to the chemical potential in the grand canonical system (see Appendix A for the details).
Equations (\ref{fp}) and (\ref{stationary1}) then reads
\begin{eqnarray}
\rho^*({\bf x},t)&=&\frac{N\,e^{-\int_{{\bf x}\neq{\bf y}} d{\bf y}\beta v(|{\bf x}-{\bf y}|)\rho^*({\bf x},t)-J({\bf x})}}{\int d{\bf x}\,e^{-\int_{{\bf x}\neq{\bf y}} d{\bf y}\beta v(|{\bf x}-{\bf y}|)\rho^*({\bf x},t)-J({\bf x)}}},\nonumber\\
\beta\mu&=&\ln\left(
\frac{N}{\int d{\bf x}\,e^{-\int_{{\bf x}\neq{\bf y}} d{\bf y}\beta v(|{\bf x}-{\bf y}|)\rho^*({\bf x},t)-J({\bf x)}}}
\right).
\label{mf}
\end{eqnarray}
While eq. (\ref{mf}) has been used widely and successfully as the Poisson-Boltzmann equation when $v(|{\bf x}-{\bf y}|)$ is set to be the pure Coulomb potential $v(|{\bf x}-{\bf y}|)\sim 1/|{\bf x}-{\bf y}|$, the solution of eq. (\ref{mf}) for hard sphere fluid provides a cavity region in the particle core, indicating that eq. (\ref{mf}) is irrelevant to describe the short-range correlations on particle size scale and is relevant at a coarse-grained scale. 
A variety of self-consistent equations has therefore been developed in the framework of the liquid state theory including the DFT [1-5].
Incidentally, the DFT has been found to be among the most powerful tools to predict the density distribution in equilibrium, which is derived not from the above functional, $F[\rho]$, but from the Hohenberg-Kohn free energy functional $\mathcal{F}_{\mathrm{HK}}[\rho]$ [1-5] defined below.
Nevertheless, it will be demonstrated in the present and next sections that the density distribution $\rho^*$ at a coarse-grained scale plays a relevant role as an indicator both in the stochastic dynamics and thermodynamics around a metastable state.

In the following, we investigate the stochastic dynamics of $\phi=\rho-\rho^*$ representing the density fluctuations around $\rho^*$ satisfying eq. (\ref{stationary1}), which will be referred to as the density distribution in a metastable state.
We expand the left hand side of eq. (\ref{stationary1}) as
\begin{equation}
\frac{\delta F[\rho]}{\delta\rho({\bf x},t)}
=\left.
\frac{\delta F[\rho]}{\delta\rho({\bf x},t)}\right|_{\rho=\rho^*}
+\int d{\bf y}\left.
\frac{\delta^2 F[\rho]}{\delta\rho({\bf x},t)\delta\rho({\bf y},t)}\right|_{\rho=\rho^*}
\phi({\bf y},t),
\label{expansion}
\end{equation}
Noting that $\nabla\mu=0$, combination of eqs. (\ref{sdft0}), (\ref{stationary1}) and (\ref{expansion}) yields a set of equations:
\begin{eqnarray}
\fl
\partial_t\rho^*({\bf x},t)=\zeta[\rho^*,{\bf \eta}]=\nabla\cdot\sqrt{2D\rho^*({\bf x},t)}{\bf \eta}({\bf x},t),
\label{set1}\\
\fl
\partial_t\{\rho^*({\bf x},t)+\phi({\bf x},t)\}=\nabla\cdot D\rho^*\left(1+\frac{\phi}{\rho^*}\right)\nabla\int d{\bf y}\left.
\frac{\delta^2 (\beta F[\rho])}{\delta\rho({\bf x},t)\delta\rho({\bf y},t)}\right|_{\rho=\rho^*}
\phi({\bf y},t)
+\zeta[\rho^*\left(1+\phi/\rho^*\right),{\bf \eta}].\nonumber\\
\label{set2}
\end{eqnarray}
Equation (\ref{set1}) also reads
\begin{equation}
\fl
-\partial_t\rho^*({\bf x},t)=\partial_t\rho^*({\bf x},-t)=\nabla\cdot\sqrt{2D\rho^*({\bf x},-t)}{\bf \eta}({\bf x},-t).
\label{set3}
\end{equation}
When we consider the $\phi$-dynamics satisfying $\phi/\rho^*\ll 1$, we obtain from adding eqs. (\ref{set2}) and (\ref{set3})
\begin{eqnarray}
\partial_t\phi=\nabla\cdot D\rho^*\nabla\int d{\bf y}\left.
\frac{\delta^2 (\beta F[\rho])}{\delta\rho({\bf x},t)\delta\rho({\bf y},t)}\right|_{\rho=\rho^*}
\phi({\bf y},t)
+\sqrt{2}\zeta[\rho^*,{\bf \eta}],\nonumber\\
\label{sdft2}
\end{eqnarray}
where the prefactor $\sqrt{2}$ arises from the mean square value of $\nabla\cdot\sqrt{2D\rho^*({\bf x},t)}{\bf \eta}({\bf x},t)+\nabla\cdot\sqrt{2D\rho^*({\bf x},-t)}{\bf \eta}({\bf x},-t)$.

It is noted that eq. ({\ref{sdft2}}) is similar to the model-B equation that has been proposed in terms of the linearized stochastic DFT, but that a couple of different points exists as seen from the model-B equation [37-46]:
\begin{eqnarray}
\partial_t\phi=\nabla\cdot D\rho^*\nabla\int d{\bf y}\left.
\frac{\delta^2 (\beta \mathcal{F}_{\mathrm{HK}}[\rho])}{\delta\rho({\bf x},t)\delta\rho({\bf y},t)}\right|_{\rho=\rho^*}
\phi({\bf y},t)
+\zeta[\rho^*,{\bf \eta}],\nonumber\\
\label{sdft3}
\end{eqnarray}
where the free energy functional is replaced by the equilibrium free energy functional $\mathcal{F}_{\mathrm{HK}}[\rho]$ and the noise term is changed from $\sqrt{2}\zeta$ to $\zeta$.
We will see the relationship between eqs. (\ref{sdft2}) and  (\ref{sdft3}) after investigating the stochastic thermodynamics and giving a more tractable expression of the free energy functional $F[\rho]$ evaluated in the saddle-point (SP) approximation.

\section{Stochastic thermodynamics around a metastable state}

Following the previous study on the stochastic thermodynamics of systems that are described by a time-dependent density field, it is possible to investigate the change of the free energy functional due to the density fluctuation around a metastable state defined by eq. (\ref{stationary1}).
We do not only consider a process with identical initial and final states (i.e., $\rho^*({\bf x},\tau)=\rho^*({\bf x},0)$), but also fix external parameters during the process.
It is to be noted that the density difference $\rho({\bf x},t)-\rho^*({\bf x},t)=\phi({\bf x},t)$ exists for $0<t<\tau$: density fluctuations are allowed for $0<t<\tau$.
Focusing on the non-local interaction contribution in terms of the free energy functional $F[\rho]$, we have in the first approximation [33]
\begin{eqnarray}
\frac{dF[\rho]}{dt}=\int d{\bf x}
\frac{\delta F[\rho]}{\delta\rho({\bf x},t)}
\partial_t\rho({\bf x},t)
=\frac{dQ_0}{dt},
\label{q-rate}
\end{eqnarray}
where no work is done because of the invariance of external parameters, and therefore $dQ_0/dt$ denotes the heat flow dissipated into the reservoir due to the change of the free energy functional.
What is meant by the first approximation regarding eq. (\ref{q-rate}) is the neglect of the additional term,  $(1/2)\int\int d{\bf x}d{\bf y}\,\delta^2 F[\rho]/\{\delta\rho({\bf x},t)\delta\rho({\bf y},t)\}\zeta({\bf x},t)\zeta({\bf y},t)$, due to the Ito calculus when considering density fluctuations [38], which remains to be treated cautiously.

The total heat flow dissipated into the reservoir is given by $dQ/dt=dQ_0/dt+T(dS[\rho]/dt)$ [33], indicating that the constrained intrinsic entropy $S[\rho]$ associated with the set of different microstates is not considered in eq. (\ref{q-rate}).
However, the entropy as a thermodynamic function validates the following integral:
\begin{equation}
T\int_0^{\tau}dt\,\frac{dS[\rho]}{dt}=T\int_0^{\tau}dS[\rho]=S[\rho^*({\bf x},\tau)]-S[\rho^*({\bf x},0)],
\label{entropy}
\end{equation}
which vanishes in the case of our process that $\rho^*({\bf x},\tau)=\rho^*({\bf x},0)$.
Combining eqs. (\ref{expansion}) and (\ref{q-rate}), on the other hand, we have
\begin{eqnarray}
Q_0&=&\int d{\bf x}\int_0^{\tau}dt\frac{\delta F[\rho]}{\delta\rho({\bf x},t)}\partial_t\rho({\bf x},t)\nonumber\\
&=&\int d{\bf x}\int d{\bf y}\left.
\frac{\delta^2 F[\rho]}{\delta\rho({\bf x},t)\delta\rho({\bf y},t)}\right|_{\rho=\rho^*}\,q_{\tau},\nonumber\\
q_{\tau}&=&\int_0^{\tau}dt\,\phi({\bf y},t)\partial_t\rho({\bf x},t),
\label{Q}
\end{eqnarray}
where the time integration of $q_{\tau}$ by parts provides
\begin{eqnarray}
\fl
q_{\tau}=\left[\phi({\bf y},t)
\rho({\bf x},t)\right]_0^{\tau}
-\int_0^{\tau}dt\,\partial_t\phi({\bf y},t)\rho({\bf x},t)\nonumber\\
\fl
=-\int_0^{\tau}dt\,\rho({\bf x},t)
\left[
\nabla_{{\bf y}}\cdot D \rho^*\nabla_{{\bf y}}\int d{\bf z}\left.
\frac{\delta^2 (\beta F[\rho])}{\delta\rho({\bf y},t)\delta\rho({\bf z},t)}\right|_{\rho=\rho^*}
\phi({\bf z},t)
+\sqrt{2}\zeta[\rho^*({\bf y},t),{\bf \eta}({\bf y},t)]
\right],\nonumber\\
\label{q-parts}
\end{eqnarray}
using eq. (\ref{sdft2}).
Reminding that $\rho=\rho^*+\phi$ in eq. (\ref{q-parts}), we assume that the process time $\tau$ is long enough to validate $\int_0^{\tau}dt\,\phi({\bf z},t)=\int_0^{\tau}\zeta[\rho^*,{\bf \eta}]=0$.
Equation (\ref{q-parts}) thus reads that
\begin{eqnarray}
q_{\tau}=
-\left[
\nabla_{{\bf y}}\cdot D \rho^*\nabla_{{\bf y}}\int d{\bf z}\left.
\frac{\delta^2 (\beta F[\rho])}{\delta\rho({\bf y},t)\delta\rho({\bf z},t)}\right|_{\rho=\rho^*}
\right]
\int_0^{\tau}dt\phi({\bf x},t)\phi({\bf z},t).\,\nonumber\\
\label{q-parts2}
\end{eqnarray}
Furthermore, it is found that [37]
\begin{eqnarray}
\phi({\bf x},t)\phi({\bf z},t)=k_BT
\left.\left(
\frac{\delta^2 F[\rho]}{\delta\rho({\bf x},t)\delta\rho({\bf z},t)}
\right)^{-1}\right|_{\rho=\rho^*},
\label{static-corr}
\end{eqnarray}
following the previous result of the stochastic DFT.
Inserting eq. (\ref{static-corr}) into eq. (\ref{q-parts2}), eq. (\ref{Q}) is transformed to
\begin{eqnarray}
Q_0&=&
-\int d{\bf x}\int d{\bf y}\left.
\frac{\delta^2 F[\rho]}{\delta\rho({\bf x},t)\delta\rho({\bf y},t)}\right|_{\rho=\rho^*}
\nabla_{{\bf y}}\cdot D \rho^*({\bf y},t)\nabla_{{\bf y}}\,
\delta({\bf x}-{\bf y})\,\tau
\nonumber\\
&=&0.
\label{heat-result}
\end{eqnarray}
Equations (\ref{entropy}) and (\ref{heat-result}) imply that the heat dissipated into the reservoir is negligible on average in the first approximation, even when small fluctuations around the metastable density distribution determined by eq. (\ref{stationary1}) are considered during the process.

\section{The free energy functional $F[\rho]$ of a given density profile in the saddle-point (SP) approximation}

We make use of the formulation of the DFT for obtaining an approximate form of $F[\rho]$ given by eq. (8).
As detailed in Appendix A, the Fourier transform representation of the distribution functional, 
$P_{\mathrm{eq}}[\rho,t]=\left<\prod_{{\bf x}}\delta\left[
\rho({\bf x},t)-\hat{\rho}({\bf x},t)
\right]\right>_{\mathrm{eq}}$,
in equilibrium is given by
\begin{eqnarray}
\fl
P_{\mathrm{eq}}[\rho,t]
=\frac{1}{e^{-\beta\Omega[J]}}
\int \mathcal{D}\psi\,
\mathrm{Tr}\,
\,e^{-\frac{1}{2}\int d{\bf x}\int d{\bf y}\,
\mathcal{G}_{\hat{\rho}}({\bf x},{\bf y})\beta v(|{\bf x}-{\bf y}|)
-\int d{\bf x}i\hat{\rho}({\bf x},t)\psi({\bf x})}
e^{\int d{\bf x}\,\rho({\bf x},t)\{i\psi({\bf x})-J({\bf x})\}}\nonumber\\
\fl\qquad\quad
=\frac{1}{e^{-\beta\Omega[J]}}
\int \mathcal{D}\psi\,
e^{-\beta\Omega[i\psi]+\int d{\bf x}\,\rho({\bf x},t)\{i\psi({\bf x})-J({\bf x})\}},
\label{given-free1}
\end{eqnarray}
with $\Omega$ denoting the grand potential.
It is to be noted that the above expression (\ref{given-free1}) provides the free energy functional $F[\rho]$ of a given density different from the equilibrium density, albeit using the equilibrium grand potential $\Omega$.
In what follows, we show that the appearance of $\Omega$ in eq. (\ref{given-free1}) allows to evaluate the difference between $F[\rho]$ and the equilibrium free energy, denoted as $\mathcal{F}_{\mathrm{HK}}[\rho]+k_BT\int d{\bf x}\rho({\bf x},t)J({\bf x})$ below, where $\rho$ is identified with the equilibrium distribution.

We evaluate the functional integral over the $\psi$-field in the SP approximation that includes the Gaussian fluctuations around the SP field $\psi=i\psi^*$ determined by the SP equation as follows:
\begin{equation}
\left.
\frac{\delta\left(\beta\Omega[i\psi]\right)}{\delta\psi({\bf x})}\right|_{\psi=i\psi^*}
=i\rho({\bf x},t),
\label{sp}
\end{equation}
which reads
\begin{equation}
\left<\hat{\rho}({\bf x},t)\right>_{\mathrm{eq}}
=\rho({\bf x},t),
\end{equation}
using the relation:
\begin{eqnarray}
\fl
\left.
\frac{\delta\left(\beta\Omega[i\psi]\right)}{\delta\psi({\bf x})}\right|_{\psi=i\psi^*}
&=&-\frac{\delta}{\delta\psi({\bf x})}
\ln
\left.\left[
\mathrm{Tr}\,
\,e^{-\frac{1}{2}\int d{\bf x}\int d{\bf y}\,
\mathcal{G}_{\hat{\rho}}({\bf x},{\bf y})\beta v(|{\bf x}-{\bf y}|)
-\int d{\bf x}i\hat{\rho}({\bf x},t)\psi({\bf x})}
\right]
\right|_{\psi=i\psi^*}\nonumber\\
&=&i\left<\hat{\rho}({\bf x},t)\right>_{\mathrm{eq}}.
\end{eqnarray}
The SP equation (\ref{sp}) is similar to one of the significant equations for determining the equilibrium density distribution in the DFT [1-5].
Actually, the first Legendre transform of $\Omega[i\psi=-\psi^*]$ defines the free energy functional that can be identified with the Hohenberg-Kohn free energy, the central functional of the DFT:
\begin{equation}
\beta\mathcal{F}_{\mathrm{HK}}[\rho]
=\beta\Omega[-\psi^*]+\int d{\bf x}\rho({\bf x},t)\psi^*({\bf x}),
\label{hk}
\end{equation}
which also satisfies the following identity as well as that in the DFT:
\begin{eqnarray}
\frac{\delta\left(\beta\mathcal{F}_{\mathrm{HK}}[\rho]\right)}{\delta\rho({\bf x})}
=\psi^{*}({\bf x}).
\label{dft}
\end{eqnarray}
Setting that $\psi^*=\mu-J$ in eq. (\ref{dft}), a given density of $\rho$ represents in itself the density distribution in equilibrium as proved by the DFT. 

The local minimum and maximum are located at the SP field, $i\psi^*({\bf x})$, along the real and imaginary fields, respectively.
Due to the functional configuration around the SP field, what is relevant to the Gaussian approximation is a quadratic term with respect to a fluctuating field, $\Delta\psi_R$, along the real axis.
It follows that
\begin{equation}
\fl
\beta\Omega[i\psi]-\int d{\bf x}i\rho({\bf x},t)\psi({\bf x})
=\beta\mathcal{F}_{\mathrm{HK}}[\rho]+\frac{1}{2}\int d{\bf x}\int d{\bf y}\,
\left.
\frac{\delta^2\left(\beta\Omega[i\psi]\right)}{\delta\psi({\bf x})\delta\psi({\bf y})}
\right|_{\psi=i\psi^*}\Delta\psi_R({\bf x})\Delta\psi_R({\bf y}),
\label{omega-expansion}
\end{equation}
where the second term on the right hand side (rhs) of eq. (\ref{omega-expansion}) yields the additional free energy functional that is obtained from the Gaussian integration over the $\Delta\psi_R$-field:
\begin{eqnarray}
\fl
e^{-\beta\Psi}
=\int \mathcal{D}{\Delta\psi}_R
\exp\left[
-\frac{1}{2}\int d{\bf x}\int d{\bf y}\,
\left.
\frac{\delta^2\left(\beta\Omega[i\psi]\right)}{\delta\psi({\bf x})\delta\psi({\bf y})}
\right|_{\psi=i\psi^*}\Delta\psi_R({\bf x})\Delta\psi_R({\bf y})
\right].
\label{psi1}
\end{eqnarray}
As shown in Appendix B, $\beta\Psi$ is written as
\begin{eqnarray}
\beta\Psi[\rho]=\frac{1}{2}\ln\left|
\mathrm{det}\left\{
\delta({\bf x}-{\bf y})+\rho({\bf y})h(|{\bf x}-{\bf y}|)\right\}
\right|,
\label{psi2}
\end{eqnarray}
where the pair correlation function $h(|{\bf x}-{\bf y}|)$ is related to the direct correlation function $c(|{\bf x}-{\bf y}|)$ via the Orstein-Zernike equation [1]:
\begin{eqnarray}
h(|{\bf x}-{\bf y}|)=c(|{\bf x}-{\bf y}|)+\int d{\bf z}\,
\rho({\bf z})\,h(|{\bf x}-{\bf z}|)c(|{\bf z}-{\bf y}|).
\label{oz}
\end{eqnarray}
Going back to eq. (\ref{given-free1}), we thus obtain an approximate form of $F[\rho]$:
\begin{equation}
\beta F[\rho]=\beta \mathcal{F}_{\mathrm{HK}}[\rho]+\beta\Psi[\rho]+\int d{\bf x}\rho({\bf x},t)J({\bf x}),
\label{f-sum}
\end{equation}
where $\mathcal{F}_{\mathrm{HK}}[\rho]$ and $\beta \Psi[\rho]$ are expressed as eqs. (\ref{hk}) and (\ref{psi2}), respectively.
Equation (\ref{f-sum}) reveals that $\Psi[\rho]$ given by eq. (\ref{psi2}) is an additional term to the equilibrium free energy, $\mathcal{F}_{\mathrm{HK}}[\rho]+k_BT\int d{\bf x}\rho({\bf x},t)J({\bf x})$, when $\rho$ is different from the equilibrium density.

\section{Stationary equation of $F[\rho]$ in a metastable state}

Combining eqs. (\ref{stationary1}), (\ref{dft}) and (\ref{f-sum}), we have
\begin{eqnarray}
\left.
\frac{\delta\left(\beta F[\rho]\right)}{\delta\rho({\bf x})}
\right|_{\rho=\rho^*}=\psi^{*}({\bf x})
+\left.
\frac{\delta\left(\beta \Psi[\rho]\right)}{\delta\rho({\bf x})}
\right|_{\rho=\rho^*}
+J({\bf x})=\mu.
\label{stationary2}
\end{eqnarray}
As found from Appendix B, the contribution of $\Psi$, or the second term on the rhs of eq. (\ref{stationary2}),  is simply given by
\begin{equation}
\left.
\frac{\delta\left(\beta \Psi[\rho]\right)}{\delta\rho({\bf x})}
\right|_{\rho=\rho^*}
=\frac{\rho^*({\bf x})c(0)-1}{2},
\label{psi}
\end{equation}
in the Percus-Yevick (PY) approximation [1, 47] of hard sphere fluids.
If we can ignore the above contribution (eq. (\ref{psi})) arising from $\Psi$, eqs. (\ref{dft}) and (\ref{stationary2}) may be reduced to the stationary equation with respect to the equilibrium density $\rho_{\mathrm{eq}}$:
\begin{eqnarray}
\mu-J({\bf x})
=\left.
\frac{\delta(\beta \mathcal{F}_{\mathrm{HK}}[\rho])}{\delta\rho({\bf x},t)}\right|_{\rho=\rho_{\mathrm{eq}}}
=\psi^*({\bf x}).
\label{eqrho}
\end{eqnarray}
Similarly, eq. (\ref{stationary2}) reads in the PY approximation:
\begin{eqnarray}
\mu-J({\bf x})
=\left.
\frac{\delta(\beta \mathcal{F}_{\mathrm{HK}}[\rho])}{\delta\rho({\bf x},t)}\right|_{\rho=\rho^*}
+\frac{\rho^*({\bf x})c(0)-1}{2},
\label{metarho}
\end{eqnarray}
using eq. (\ref{psi}).
It is found from comparing eqs. (\ref{eqrho}) and (\ref{metarho}) that the SP potential field $\delta(\beta \mathcal{F}_{\mathrm{HK}}[\rho])/\delta\rho=\psi^*$ varies because the density is changed from $\rho_{\mathrm{eq}}$ to $\rho^*$.
We would also like to note that minus the direct correlation function of a repulsive interaction is larger at ${\bf x}={\bf y} $ as the density is higher: $1-\rho^*({\bf x})c(0)>0$.
Equations (\ref{eqrho}) and (\ref{metarho}) therefore imply that $\delta\mathcal{F}_{\mathrm{HK}}/\delta\rho|_{\rho=\rho^*}>\delta\mathcal{F}_{\mathrm{HK}}/\delta\rho|_{\rho=\rho_{\mathrm{eq}}}$, or $\rho^*>\rho_{\mathrm{eq}}$ in the PY fluids.
In other words, the metastable density should be higher than the equilibrium one in the case of the PY fluids at the same condition such as identical temperature and chemical potential, sharing a property of supercooled liquids.

\section{Stochastic DFT around a metastable state using an approximate form of $F[\rho]$}

It is found from eqs. (\ref{sdft2}) and (\ref{f-sum}) that the stochastic DFT further reads
\begin{eqnarray}
\fl
\partial_t\phi=\nabla\cdot D\rho^*\nabla\int d{\bf y}
\left[\left.
\frac{\delta^2 \mathcal{F}_{\mathrm{HK}}[\rho]}{\delta\rho({\bf x},t)\delta\rho({\bf y},t)}\right|_{\rho=\rho^*}
+\left.
\frac{\delta^2\left(\beta\Psi\right)}{\delta\rho({\bf x},t)\delta\rho({\bf x},t)}
\right|_{\rho=\rho^*}
\right]
\phi({\bf y},t)
+\sqrt{2}\zeta[\rho^*,{\bf \eta}],\nonumber\\
\label{sdft-bare}
\end{eqnarray}
where it is well known that [1-5]
\begin{eqnarray}
\left.
\frac{\delta^2 \mathcal{F}_{\mathrm{HK}}[\rho]}{\delta\rho({\bf x},t)\delta\rho({\bf y},t)}\right|_{\rho=\rho^*}
=\frac{\delta({\bf x}-{\bf y})}{\rho^*({\bf x})}-c(|{\bf x}-{\bf y}|).
\label{second-hk}
\end{eqnarray}
The additional term due to $\Psi$ is evaluated in Appendix B, providing in the PY approximation of hard sphere fluids:
\begin{eqnarray}
\left.
\frac{\delta^2\left(\beta\Psi\right)}{\delta\rho({\bf x},t)\delta\rho({\bf x},t)}
\right|_{\rho=\rho^*}
=\frac{c(0)}{2}+\frac{\rho({\bf x})}{2}\left(\frac{\delta c(0)}{\delta\rho^*}\right).
\label{psi-second}
\end{eqnarray}
Accordingly, we have
\begin{eqnarray}
\fl
\partial_t\phi=
\nabla\cdot D\rho^*\nabla\,\chi[\rho^*]\phi({\bf x},t)
-\nabla\cdot D\rho^*\nabla\int_{{\bf x}\neq{\bf y}} d{\bf y}
c(|{\bf x}-{\bf y}|)\phi({\bf y},t)
+\sqrt{2}\zeta[\rho^*,{\bf \eta}],\nonumber\\
\chi[\rho^*]=\frac{1}{\rho^*({\bf x})}-\frac{c(0)}{2}+\frac{\rho({\bf x})}{2}\left(\frac{\delta c(0)}{\delta\rho^*}\right),
\label{sdft-ln}
\end{eqnarray}
dividing the rhs of the above equation into the local and non-local contributions.

In the PY approximation of hard sphere fluids uniformly distributed, $-c(0)$ is identified with the compressibility [47]:
\begin{equation}
-c(0)=\beta\frac{\partial P}{\partial\overline{\rho}},
\label{cmp}
\end{equation}
where $\overline{\rho}$ denotes the smeared density.
We can investigate the difference between eqs. (\ref{sdft2}) and (\ref{sdft3}) using the expressions (\ref{sdft-ln}) and (\ref{cmp}) of the PY fluids.
For the stochastic dynamics in equilibrium of the PY fluids, eq. (\ref{sdft3}) is reduced to
\begin{eqnarray}
\fl
\partial_t\phi=
\left\{
1+\beta\overline{\rho}
\frac{\partial P}{\partial\overline{\rho}}
\right\}D
\Delta \phi({\bf x},t)
-\nabla\cdot D\overline{\rho}\nabla\int_{{\bf x}\neq{\bf y}} d{\bf y}
c(|{\bf x}-{\bf y}|)\phi({\bf y},t)
+\zeta[\rho^*,{\bf \eta}],
\label{py-eq}
\end{eqnarray}
where $\rho^*({\bf x},t)$ in eq. (\ref{sdft3}) is approximated by $\overline{\rho}$.
For the stochastic dynamics around a metastable state of the PY fluids, on the other hand, eqs. (\ref{sdft-ln}) and (\ref{cmp})  transform eq. (\ref{sdft2}) to
\begin{eqnarray}
\fl
\partial_t\phi=
\left\{
1+\frac{\beta\overline{\rho}}{2}\left\{
\frac{\partial P}{\partial\overline{\rho}}\right\}
-\frac{\beta\overline{\rho}^2}{2}\left\{
\frac{\partial^2 P}{\partial\overline{\rho}^2}\right\}
\right\}
D
\Delta \phi({\bf x},t)\nonumber\\
\fl\hspace{48mm}
-\nabla\cdot D\overline{\rho}\nabla\int_{{\bf x}\neq{\bf y}} d{\bf y}
c(|{\bf x}-{\bf y}|)\phi({\bf y},t)
+\sqrt{2}\zeta[\rho^*,{\bf \eta}].
\label{py-meta}
\end{eqnarray}
It is confirmed using the analytical expression [1, 47] of $c(0)$ in the PY approximation that $\partial^2 P/\partial\overline{\rho}^2$ is positive and is an increasing function of $\overline{\rho}$.
Equations (\ref{py-eq}) and (\ref{py-meta}) therefore indicate that the local contribution to the effective diffusion constant is decreased due to the stochastic dynamics around a metastable state instead of the equilibrium state, and that the suppression effect on the particle diffusion is enhanced with the increase of $\overline{\rho}$ in the case of the PY fluids.

\section{Concluding remarks}

To summarize, we have developed a linear stochastic DFT that applies to the stochastic dynamics and thermodynamics around a metastable state, which is essentially due to the static results of eqs. (\ref{stationary1}) and (\ref{f-sum}).
It is clear why we focus on the stationary condition (\ref{stationary1}) regarding the free energy functional $F[\rho]$ of a given density profile:
the DK equation (\ref{sdft0}) can be linearized in a straightforward manner, by considering small fluctuations around the density distribution that satisfies eq. (\ref{stationary1}).
Moreover, our field theoretic treatment newly developed herein leads to eq.  (\ref{f-sum}), revealing that $F[\rho]$ can be approximately expressed by the sum of the equilibrium free energy functional $\mathcal{F}_{\mathrm{HK}}$ used in the equilibrium DFT and the correction term $\Psi$.
Equations (\ref{stationary1}) and (\ref{f-sum}) allow the DK equation (\ref{sdft0}) to be transformed to eq. (\ref{sdft-ln}), which is our main result.

We would like to make a few remarks on the interpretation of the present metastability determined by eq. (\ref{stationary1}).
It is to be noted that the metastable density distribution $\rho^*$ obeys the stochastic equation given by eq. (\ref{set1}), implying that $\rho^*$ has a kind of the Goldstone mode.
The details follow:
The $\rho^*$-dynamics has no characteristic time of fluctuations due to the disappearance of the first two terms on the rhs of the stochastic density functional equation (\ref{sdft-ln}), while satisfying one of the solutions to eq. (\ref{stationary1}).
The noise term would cause global transitions among a set of the translationally invariant solutions to eq. (\ref{stationary1}), if exists.
Driven by the Goldstone-like dynamics, particles travel to explore the truly equilibrium state, which is meant by the present metastability.
Actually, we ignore the dissipation due to the Goldstone mode in section 4.
That is, the result (\ref{heat-result}) has been obtained under the assumption that the initial and final states are identical (i.e., $\rho^*({\bf x},\tau)=\rho^*({\bf x},0)$).

Our stochastic DFT remains to be validated more quantitatively.
The PY fluids described by eq. (\ref{py-meta}) would be available for this purpose not only because the analytical form of the direct correlation function is given [47], but also because of the clear physical meaning of $\alpha(\overline{\rho})$ in eq. (\ref{py-meta}).
It is possible to evaluate the metastable density using eq. (\ref{metarho}) and to compare the effective diffusion constants obtained from eqs. (\ref{py-eq}) and (\ref{py-meta}) in the PY approximation, which needs to be addressed in the future.

\appendix
\section{Derivation of eq. (\ref{peq-answer}) using the grand canonical ensemble}

Here we derive eq. (\ref{peq-answer}) via the formulations in the grand canonical system [20, 48].
Let us first define the grand canonical distribution functional $P^G_{\mathrm{eq}}[\rho,t]$ of a given density profile as follows:
\begin{eqnarray}
\fl
P^G_{\mathrm{eq}}[\rho,t]=\frac{1}{e^{-\beta\Omega[J]}}
\mathrm{Tr}\,
\prod_{{\bf x}}\delta\left[
\rho({\bf x},t)-\hat{\rho}({\bf x},t)
\right]
\,e^{-\frac{1}{2}\int d{\bf x}\int d{\bf y}\,
\mathcal{G}_{\hat{\rho}}({\bf x},{\bf y})\beta v(|{\bf x}-{\bf y}|)
-\int d{\bf x}\hat{\rho}({\bf x},t)J({\bf x})},\nonumber\\
\label{peqg}
\end{eqnarray}
where $\Omega[J]$ denotes the grand potential given by
\begin{eqnarray}
\fl
e^{-\beta\Omega[J]}=\mathrm{Tr}\,\exp\left[-\frac{1}{2}\int d{\bf x}\int d{\bf y}\,
\mathcal{G}_{\hat{\rho}}({\bf x},{\bf y})\beta v(|{\bf x}-{\bf y}|)
-\int d{\bf x}\hat{\rho}({\bf x},t)J({\bf x})
\right],\nonumber\\
\fl
\mathrm{Tr}\equiv
\sum_{N=0}^{\infty}e^{N\beta\mu}\mathrm{Tr}_N,
\label{f-start}
\end{eqnarray}
with introducing the configurational integrals in the canonical system: $\mathrm{Tr}_N\equiv(1/N!)\int d{\bf x}_1\cdots\int d{\bf x}_N$.
The Fourier transform of the delta functional is written as
\begin{equation}
\prod_{{\bf x}}\delta\left[
\rho({\bf x},t)-\hat{\rho}({\bf x},t)
\right]
=\int \mathcal{D}\psi\,e^{\int d{\bf x}\,i\psi({\bf x},t)\{\rho({\bf x},t)-\hat{\rho}({\bf x},t)\}},
\end{equation}
which is plugged into eq. (\ref{peqg}) to obtain the following form:
\begin{eqnarray}
\fl
P^G_{\mathrm{eq}}[\rho,t]=\frac{1}{e^{-\beta\Omega[J]}}
\int \mathcal{D}\psi\,
\mathrm{Tr}\,
e^{-\int d{\bf x}\,i\psi({\bf x})\hat{\rho}({\bf x},t)}
\nonumber\\
\fl\qquad\qquad
\times
e^{-\frac{1}{2}\int d{\bf x}\int d{\bf y}\,
\mathcal{G}_\rho({\bf x},{\bf y})\beta v(|{\bf x}-{\bf y}|)
-\int d{\bf x}\rho({\bf x},t)J({\bf x})+\int d{\bf x}\,i\psi({\bf x})\rho({\bf x},t)},\nonumber\\
\fl
\mathrm{Tr}\,
e^{-\int d{\bf x}\,i\psi({\bf x})\hat{\rho}({\bf x},t)}=\exp\left\{
\int d{\bf x}\,e^{\beta\mu-i\psi({\bf x})}
\right\}.
\label{peqg2}
\end{eqnarray}
As shown in the previous study, the saddle-point approximation of the $\psi$-field in eq. (\ref{peqg2}) yields
\begin{eqnarray}
\fl
P^G_{\mathrm{eq}}[\rho,t]=\frac{1}{e^{-\beta\Omega[J]}}
\exp\left\{
-\beta F[\rho]+\int d{\bf x}\,\beta\mu\rho({\bf x},t)-\frac{1}{2}\ln\mathrm{det}\rho
\right\},
\label{peqg3}
\end{eqnarray}
where the last term in the exponent on the rhs arises from the Gaussian integration over the fluctuating field around the SP field $\psi^*$.
We can go back from the configurational integrals, denoted by $\mathrm{Tr}$, in the grand canonical system to $\mathrm{Tr}_N$ in the canonical system via the contour integral such that
\begin{eqnarray}
\fl
\mathrm{Tr}\,
\prod_{{\bf x}}\delta\left[
\rho({\bf x},t)-\hat{\rho}({\bf x},t)
\right]
\,e^{-\frac{1}{2}\int d{\bf x}\int d{\bf y}\,
\mathcal{G}_{\hat{\rho}}({\bf x},{\bf y})\beta v(|{\bf x}-{\bf y}|)
-\int d{\bf x}\hat{\rho}({\bf x},t)J({\bf x})}\nonumber\\
\fl\qquad
=\frac{1}{2\pi i}\oint dz\,\frac{e^{-\beta\Omega[J]}P^G_{\mathrm{eq}}[\rho,t]}{z^{N+1}}
=\frac{1}{2\pi i}\oint dz\,\frac{e^{-\beta F[\rho]-\frac{1}{2}\ln\det\rho}}{z^{N-\int d{\bf x}\,\rho({\bf x},t)+1}},
\label{contour}
\end{eqnarray}
where $z=e^{\beta\mu}$ is now a complex variable.
Combining eqs. (\ref{peqg}), (\ref{f-start}) and (\ref{contour}), we have
\begin{eqnarray}
P_{\mathrm{eq}}[\rho,t]
=P^G_{\mathrm{eq}}[\rho,t]\delta\left(
\int d{\bf x}\,\rho({\bf x},t)-N
\right).
\end{eqnarray}
Furthermore, we evaluate the correction term $(1/2)\ln(\mathrm{det}\rho)$ written as
\begin{eqnarray}
\frac{1}{2}\ln\mathrm{det}\rho&=&\int d{\bf x}\lim_{a\rightarrow 0}\frac{1}{2a^3}\ln(\rho_la^3),\nonumber\\
\lim_{a\rightarrow 0}\frac{1}{a^3}\ln(\rho_la^3)&=&\lim_{a\rightarrow 0}\left(\rho_l-\overline{\rho}+\mathcal{O}[a^3]\right)
=\rho({\bf x})-\overline{\rho},
\label{log}
\end{eqnarray}
where $a$ denotes the lattice constant defined by $\overline{\rho}a^3=1$, we have used the discretized expression of the Gaussian integration while taking the continuum limit of $a\rightarrow 0$, and the logarithmic expansion have been performed as $\ln(\rho_la^3)=\ln\{1+(\rho_l-\overline{\rho})a^3\}\approx(\rho_l-\overline{\rho})a^3+\mathcal{O}[\{(\rho_i-\overline{\rho})a^3\}^2]$.
It follows from eq. (\ref{log}) that the correction term, $(1/2)\ln(\mathrm{det}\rho)$, is negligible:
$(1/2)\ln(\mathrm{det}\rho)=0.5(N-N)=0$ [48].
We have thus verified eq. (\ref{peq-answer}) given by
\begin{eqnarray}
P_{\mathrm{eq}}[\rho,t]
&=&\left<\prod_{{\bf x}}\delta\left[
\rho({\bf x},t)-\hat{\rho}({\bf x},t)
\right]\right>_{\mathrm{eq}}\nonumber\\
&=&\frac{e^{-\beta F[\rho]}}{\int \mathcal{D}\rho\,e^{-\beta F[\rho]}}\delta\left(
\int d{\bf x}\,\rho({\bf x},t)-N
\right),
\end{eqnarray}
using the functional integral measure $\mathcal{D}\rho$ that satisfies $\int d{\bf x}\,\rho({\bf x},t)=N$;
in the absence of the number constraint, the denominator becomes equal to the grand potential: $e^{-\beta\Omega[J]}=\int \mathcal{D}\rho\,e^{-\beta F[\rho]}$.

\section{Derivation of eqs. (\ref{psi2}), (\ref{psi}) and (\ref{psi-second})}

As found from eq. (\ref{f-sum}), the additional free energy $\beta\Psi[\rho]$ represents the difference between the free energy functionals in equilibrium and in a metastable state, and it is necessary for evaluating $\beta\Psi[\rho]$ given by eq. (\ref{psi1}) to write as
\begin{eqnarray}
\fl\left.
\frac{\delta^2\left(\beta\Omega[i\psi]\right)}{\delta\psi({\bf x})\delta\psi({\bf y})}
\right|_{\psi=i\psi^*}
&=&\left<\hat{\rho}({\bf x},t)\hat{\rho}({\bf y},t)\right>_{\mathrm{eq}}
-\left<\hat{\rho}({\bf x},t)\right>_{\mathrm{eq}}\left<\hat{\rho}({\bf y},t)\right>_{\mathrm{eq}}\nonumber\\
&=& \left<\mathcal{G}_{\hat{\rho}}\right>_{\mathrm{eq}} -\rho({\bf x},t)\rho({\bf y},t)
+\rho({\bf x},t)\delta(|{\bf x}-{\bf y}|)\nonumber\\
&=&\rho({\bf x},t)\left\{\delta({\bf x}-{\bf y})+\rho({\bf y},t)h(|{\bf x}-{\bf y}|)\right\},
\end{eqnarray}
where the pair correlation function $h(|{\bf x}-{\bf y}|)$ is defined by $\left<\mathcal{G}_{\hat{\rho}}\right>_{\mathrm{eq}} -\rho({\bf x},t)\rho({\bf y},t)\equiv\rho({\bf x},t)\rho({\bf y},t)h(|{\bf x}-{\bf y}|)$ and is related to the direct correlation function $c(|{\bf x}-{\bf y}|)$ via the Orstein-Zernike equation [1]:
\begin{eqnarray}
h(|{\bf x}-{\bf y}|)=c(|{\bf x}-{\bf y}|)+\int d{\bf z}\,
\rho({\bf z})\,h(|{\bf x}-{\bf z}|)c(|{\bf z}-{\bf y}|),
\label{app-oz}
\end{eqnarray}
which is equivalent to the following relation [1-5]:
\begin{eqnarray}
-\int d{\bf z}
\frac{\delta^2\left(\beta\mathcal{F}_{\mathrm{HK}}[\rho]\right)}{\delta\rho({\bf x})\delta\rho({\bf z})}
\frac{\delta^2\left(\beta\Omega[-\psi^*]\right)}{\delta\psi^*({\bf z})\delta\psi^*({\bf y})}
=\delta({\bf x}-{\bf y}).
\end{eqnarray}
Accordingly, eq. (\ref{psi1}) reads
\begin{eqnarray}
\beta\Psi[\rho]&=&\frac{1}{2}\ln\mathrm{det}\rho({\bf x})
+\frac{1}{2}\ln\left|\mathrm{det}\left\{
\delta({\bf x}-{\bf y})+\rho({\bf y})h(|{\bf x}-{\bf y}|)\right\}
\right|\nonumber\\
&=&\frac{1}{2}\ln\left|
\mathrm{det}\left\{
\delta({\bf x}-{\bf y})+\rho({\bf y})h(|{\bf x}-{\bf y}|)\right\}
\right|.
\label{app-psi-ln}
\end{eqnarray}
In the second line of the above equation, the evaluation (\ref{log}) given in Appendix A has been used.
Combining eqs. (\ref{app-oz}) and (\ref{app-psi-ln}), we have
\begin{eqnarray}
\fl\left.
\frac{\delta\left(\beta\Psi\right)}{\delta\rho({\bf x},t)}
\right|_{\rho=\rho^*}
=\frac{1}{2}\int d{\bf y}\,\left\{
\delta({\bf x}-{\bf y})-\rho^*({\bf x})c(|{\bf x}-{\bf y}|)\right\}
\left.
\frac{\delta \left\{\rho({\bf y})h(|{\bf x}-{\bf y}|)\right\}}{\delta\rho({\bf x})}\right|_{\rho=\rho^*}
\nonumber\\
\fl
=\frac{1}{2}\int d{\bf y}\,
\left\{
\delta({\bf x}-{\bf y})-\rho^*({\bf x})c(|{\bf x}-{\bf y}|)\right\}
\left\{
\delta({\bf x}-{\bf y})h(|{\bf x}-{\bf y}|)+
\rho^*({\bf y})\frac{\delta h(|{\bf x}-{\bf y}|)}{\delta\rho^*({\bf x})}
\right\}
\nonumber\\
\fl=
\frac{\rho^*({\bf x})c(0)-1}{2}+\frac{\rho^*({\bf x})}{2}\Delta h',
\nonumber\\
\fl
\Delta h'\equiv
\frac{\delta h(0)}{\delta\rho^*({\bf x})}
-\int d{\bf y}\,
c(|{\bf x}-{\bf y}|)\rho^*({\bf y})\frac{\delta h(|{\bf x}-{\bf y}|)}{\delta\rho^*({\bf x})}\nonumber\\
\fl\hspace{8mm}
=-\int d{\bf y}\,
c(|{\bf x}-{\bf y}|)\rho^*({\bf y})\frac{\delta h(|{\bf x}-{\bf y}|)}{\delta\rho^*({\bf x})},\nonumber\\
\label{app-first-rho}
\end{eqnarray}
where we set that $\delta h(|{\bf x}-{\bf y}|)/\delta\rho|_{\rho=\rho^*}\equiv\delta h(|{\bf x}-{\bf y}|)/\delta\rho^*$ for brevity and use has been made of the following relations: $\delta h(0)/\delta\rho^*=\delta h(|{\bf x}-{\bf y}|)/\delta\rho|_{{\bf x}={\bf y}, \rho=\rho^*}=0$ due to $h(0)=-1$ and
\begin{eqnarray}
\fl
\int d{\bf z}
\left\{
\delta({\bf x}-{\bf z})+\rho({\bf z})h(|{\bf x}-{\bf z}|)\right\}
\left\{
\delta({\bf z}-{\bf y})-\rho({\bf y})c(|{\bf z}-{\bf y}|)\right\}
=\delta({\bf x}-{\bf y}),
\label{app-oz2}
\end{eqnarray}
which is a rewrite of the Orstein-Zernike equation (\ref{app-oz}).
Equation (\ref{app-first-rho}) further yields
\begin{eqnarray}
\fl
\left.
\frac{\delta^2\left(\beta\Psi\right)}{\delta\rho({\bf x},t)\delta\rho^*({\bf x},t)}
\right|_{\rho=\rho^*}\nonumber\\
\fl
=\frac{c(0)}{2}+\frac{\rho({\bf x})}{2}\left(\frac{\delta c(0)}{\delta\rho^*}\right)
+\frac{1}{2}\Delta h'+\frac{\rho^*({\bf x})}{2}\Delta h"
-\frac{\rho^*({\bf x})}{2}
\int d{\bf y}\,
\frac{\delta \{c(|{\bf x}-{\bf y}|)\rho^*({\bf y})\}}{\delta\rho^*({\bf x})}
\frac{h(|{\bf x}-{\bf y}|)}{\delta\rho^*({\bf x})},\nonumber\\
\label{app-second-rho}
\end{eqnarray}
where $\delta c(0)/\delta\rho^*\equiv\delta c(|{\bf x}-{\bf y}|)/\delta\rho|_{{\bf x}={\bf y}, \rho=\rho^*}$ and
\begin{eqnarray}
\Delta h"\equiv
-\int d{\bf y}\,
c(|{\bf x}-{\bf y}|)\rho^*({\bf y})
\frac{\delta^2 h(|{\bf x}-{\bf y}|)}{\delta\rho^*({\bf x})^2}.
\end{eqnarray}
In the Percus-Yevick (PY) approximation of hard sphere fluids [1, 47], we have
\begin{eqnarray}
\Delta h'=\Delta h"=0,
\end{eqnarray}
and
\begin{eqnarray}
\int d{\bf y}\,
\frac{\delta \{c(|{\bf x}-{\bf y}|)\rho^*({\bf y})\}}{\delta\rho^*({\bf x})}
\frac{h(|{\bf x}-{\bf y}|)}{\delta\rho^*({\bf x})},
\end{eqnarray}
because of the integration range that is limited to the inside of hard core particle with its diameter of $d$:
it follows from $h(|{\bf x}-{\bf y}|\leq d)=-1$ that
\begin{eqnarray}
\frac{\delta h(|{\bf x}-{\bf y}|\leq d)}{\delta\rho^*({\bf x})}
=\frac{\delta^2 h(|{\bf x}-{\bf y}|\leq d)}{\delta\rho^*({\bf x})^2}
=0.
\end{eqnarray}
Accordingly, eqs. (\ref{app-first-rho}) and (\ref{app-second-rho}) are reduced to
\begin{eqnarray}
\left.
\frac{\delta\left(\beta\Psi\right)}{\delta\rho({\bf x},t)}
\right|_{\rho=\rho^*}
=
\frac{\rho^*({\bf x})c(0)-1}{2},
\label{app-first-rho2}
\end{eqnarray}
and
\begin{eqnarray}
\left.
\frac{\delta^2\left(\beta\Psi\right)}{\delta\rho({\bf x},t)\delta\rho^*({\bf x},t)}
\right|_{\rho=\rho^*}
=\frac{c(0)}{2}+\frac{\rho({\bf x})}{2}\left(\frac{\delta c(0)}{\delta\rho^*}\right),
\label{app-second-rho2}
\end{eqnarray}
respectively in the case of the PY fluids.
Equations (\ref{app-psi-ln}), (\ref{app-first-rho2}) and (\ref{app-second-rho2}) correspond to eqs. (\ref{psi2}), (\ref{psi}) and (\ref{psi-second}), respectively.


\end{document}